\documentclass[aps,prl,preprint]{revtex4}
\usepackage{epsfig}

\begin{document}

\title{The human family tree and the Neandertal branch}

\bigskip
\bigskip
\author{Maurizio Serva}
\address{Dipartimento di Matematica and I.N.F.M.
Universit\`a dell'Aquila,
I-67010 L'Aquila, Italy}
\bigskip

\date{\today}

\begin{abstract}
                     
We consider a large population of asexually reproducing 
individuals in absence of selective pressure.
The population size is maintained constant by the environment.
We find out that distances between individuals (time from the last 
common ancestor) exhibit highly non trivial properties.
In particular their distribution in a single
population is random even in the thermodynamical limit. 
As a result, not only distances are different for different 
pairs of individuals but also the mean distance of the individuals 
of a given population is different at different times.
All computed quantities are parameters free and only
scale linearly with the population size. 
Results in this paper may have some relevance in the 
'Out of Africa/ Multi-regional' debate about the origin of modern man.
In fact, the recovery of mitochondrial DNA (mtDNA) from Neandertal 
fossils in three different loci: Feldhofer (Germany), 
Mezmaiskaya (Northern Caucaso), Vinjia (Croatia),
permitted to compare Neandertal/Neandertal
distances with Neandertal/modern and modern/modern ones.

\pacs{PUT PACS HERE.000.XXX}
\end{abstract}

\maketitle

In 1997 a team of researchers~\cite{Krings1997,Krings2000}
announced that mtDNA was extracted from the humerus of the first recognized
Neandertal fossil, the individual found at the Feldhofer cave in the
Neander Valley in Germany in 1856.
In 1999 and 2000 mtDNA had been extracted from a
second Neandertal, a 29,000 year-old fossil of a baby recently discovered in
Mezmaiskaya cave in south-western Russia and  from a third Neandertal 
specimen from a cave at Vindija, Croatia~\cite{Krings2000}.

More recently,  another team~\cite{Adcock}
extracted mtDNA from a 60,000 year-old fossil of an anatomically modern human
discovered in the dry bed of Lake Mungo in New South Wales, Australia.

What makes mtDNA interestingly different from nuclear DNA is that it is
inherited only from the mother. In principle, every lineage can be
followed until the woman whose mtDNA
is the common ancestor of the mtDNA of all living humans.
This hypothetical woman is known as mitochondrial Eve.

More sketchy, we can say that mtDNA reproduces asexually
since there is not recombination as for nuclear DNA.
Therefore, assuming that the mtDNA mutates at a constant rate,
the number of differences in mtDNA between two individuals
is a measure of their distance, i.e. the number of
generations from the common ancestor.
This is particularly true for the previously
mentioned studies, since the part of mtDNA concerned
is the hypervariable region which seems to mutate
in absence of selective pressure.

On the basis of the comparison with the mtDNA of living humans, it was argued that
both Neandertals and Mungo man, should be eliminated from our ancestry. 
In fact, the distance of Neandertal from living humans was
estimated to be more then three times the average difference between 
living Sapiens or between the three Neandertals. Moreover,
Mungo man seems to carry a mtDNA which disappeared from modern humanity. 

In order to decide if these conclusions are correct
we have to understand if these differences have statistical relevance.
We address here to this problem by means of a very general model
without need to specify the details of the dynamics. 

We assume that the population size is constantly of $N$ individuals
due to ecological or environmental factors. At any generation we have 
$N$ new individuals which replace the $N$ individuals of previous generation. 
Since we deal with mtDNA, the reproduction is asexual and any individual of the 
new generation has a single parent in the previous one.
Obviously, the average number of offsprings of an individual
in the old generation is one, nevertheless, 
some of them will not have any offspring and others will have more than one.

We do not fix here the stochastic rule assigning the number of offsprings
to any individual since results do not depend on the dynamics details.
The only requirement is that the probability that two individuals in the new 
generation have the same parent is of the order of $1/N$ for large $N$.
This is a very weak and reasonable assumption:
just think at our life experience. 
To be more clear, we make two examples of stochastic rules which 
satisfy this requirement.
The first rule is that at any generation one half of the individuals 
(chosen at random) has no offsprings and the remaining part has two 
(see~\cite{Zhang}).
With this choice the probability of having the same parent for two individuals
is $1/(N-1)$, which behaves as  $1/N$ for large  $N$. 
The second rule is that any individual in the new generation chooses one
parent at random in the previous one, independently on the 
choice of the others (see~\cite{Derrida}).
In this case the probability of having the same parent for two individuals
is exactly $1/N$.

The distance between two given individuals $\alpha$ and $\beta$ 
in the same generation is, by definition, the number of 
generations from  the common ancestor. 
Since typical distances are proportional to $N$, 
as we are going to show,  it is useful to rescale them 
dividing by $N$.  Let us call $d(\alpha,\beta)$ these rescaled distances
(obviously, $d(\alpha,\alpha)$ vanishes).

For two distinct individuals $\alpha$ and $\beta$
in the same generation one has 
\begin{equation}
d(\alpha,\beta)= d(g(\alpha),g(\beta)) +1/N\;\;,
\label{distance}
\end{equation}
where $g(\alpha)$ and $g(\beta)$ are the two parent individuals.
The above  dynamics simply state that the
rescaled distance in the new generation increases by $1/N$
with respect to the parents distance
(the non rescaled distance would have increased by 1). 
The two parents coincide with probability $1/N$
and they are distinct individuals with probability $(N-1)/N$.

By using equation (1) it is easy to show that,
for $\lambda<1$, the average over the process gives
$<$$\exp(\lambda \, d(\alpha,\beta))$$>$ =$1/(1-\lambda)$.
This result, which holds for large $N$, implies that 
the probability that
$d(\alpha,\beta)=x$ is simply $\exp(-x)$.
Notice that this is not the distribution of
the distances inside a single large population but the average 
distribution sampled over many 
stochastically equivalent populations or, which is the same,
sampled over the same population at many different times.
Also notice that, according to this result,
the typical non-rescaled distance is of order $N$ (see also~\cite{Zhang}).

According to this distribution one has that the
process averages $<$$d(\alpha,\beta)$$>$=1  
and  $<$$d^2(\alpha,\beta)$$>$=2  which means that the distance between individuals 
may show huge differences for different pairs.
But this is not the important point, in fact,  there is another larger 
source of statistical dispersion for distances, as follows.

We can use again equation (1) in order to compute
the quantities $<$$d(\alpha,\beta)d(\beta,\gamma)$$>$
and  $<$$d(\alpha,\beta)d(\gamma,\delta)$$>$.
To reach this goal one simply has to take into account
that any of the pairs which can be formed by two of the four individuals
$\alpha$, $\beta$, $\gamma$ and $\delta$  may have coinciding parents 
with probability $1/N$.  The probability that
more then two parents coincide is of higher order.
Taking the large $N$ limit one finds  
3$<$$d(\alpha,\beta)d(\beta,\gamma)$$>$=$<$$d^2(\alpha,\beta)$$>$
+2$<$$d(\alpha,\beta)$$>$  and  
6$<$$d(\alpha,\beta)d(\gamma,\delta)$$>$=
4$<$$d(\alpha,\beta)d(\beta,\gamma)$$>$+
2$<$$d(\alpha,\beta)$$>$.

Solving these simple equations one gets 
$<$$d(\alpha,\beta)d(\beta,\gamma)$$>$=4/3 and
$<$$d(\alpha,\beta)d(\gamma,\delta)$$>$=11/18.
In this form, this result only seems to state that there is a 
statistical correlation between distances corresponding
to different pairs.
But there is a much more interesting consequence.
Let us introduce the mean distance of the individuals of a population as 
\begin{equation}
d= \frac{2}{N(N-1)}\sum_{\alpha>\beta} d(\alpha,\beta) \;\;,
\label{tree}
\end{equation}
this is simply the average on a single population (and at a given time)
of the internal distances considering all the $N(N-1)/2$
possible pairs. This quantity is random for finite $N$ but should
reach deterministically its average value for large $N$.

On the contrary, while the process
averages $<$$d$$>$ and  $<$$d(\alpha,\beta)$$>$  equals 1,
in the termodinamical ($N \to \infty$) limit 
$<$$d^2$$>$=$<$$d(\alpha,\beta)d(\gamma,\delta)$$>$=11/18. 
In other words, not only the distances are randomly distributed inside 
the population but their mean on all possible pairs is random
even if $N$ is extremely large. 
We show these facts in fig. 1 where the time evolution of the mean distance
of the population is shown for $N=2000$ which is
sufficiently large to destroy all finite size effects.
Looking at fig. 1, we notice that the mean distance is subject to abrupt
negative variations due to the extinction of large subpopulations. 

Notice that the typical size of living humanity, 
measured from nuclear genetic distance,
corresponds to a population of 10000 individuals.
Paleoanthropologists explain this fact 
by a recent demographic explosion which followed a bottleneck.    

Since the mean distance is random, the distribution $q(x)$
of the distances in a single large population must also be random.
For finite $N$ one has that $q(x)\, dx$ is simply the number
of pairs in a given population whose distance lies in the interval 
$[x,x+dx]$ divided by the total number $N(N-1)/2$ of possible pairs. 
Then, $d$ is simply the average on this distribution ($d=\int q(x) \, x\, dx$)
and since $d$ is random also $q(x)$ remains random when
the infinite population limit is performed.
The above definition implies $<q(x)>=p(x)$, i. e.,
the average of the $q(x)$ over many independent realizations of
the process is $\exp(-x)$. 

The most important fact is that the distribution may be very different from
the averaged one. This fact can be appreciated in fig. 2 were $q(x)$
(computed again from a population of 2000 individuals)
does not show any resemblance with its average $\exp(-x)$.
From fig. 2 (which is a typical one) it is clear that individuals
spontaneously cluster in groups.
In fact, most of the distances assume a few of values 
corresponding to the distances between the major subpopulations.  

These genetically isolated subpopulations are not different species
or geographically isolated groups, but they can originate
in perfectly inter-breeding and (nuclear DNA) homogeneous populations.
In fact, sexually reproducing nuclear DNA
has a completely different statistics.
In large populations the distance for almost all pairs of individuals
coincide with the average value 1 (see also~\cite{Serva}).

It should also noticed that distance between two individuals is estimated 
from the number of differences in mtDNA due to mutation.
The number of these differences is itself stochastic 
and only in average is proportional to the distance.
Therefore, the possibility of having large mtDNA isolation 
of  subpopulation is even larger one can estimate here.

Let us come back now to the problem which has inspired
the present work in order to have a quantitative understanding of
the phenomena.

In 1997, 1999 and 2000 a team of researchers~\cite{Krings1997,Krings2000}
extracted mtDNA from three different specimen of Neandertal and
was able to amplify many short strands of the hyper-variable region (HVR1 and HVR2)
using Polymerase Chain Reaction (PCR).
In 1997 they compared the first specimen (Feldhofer)
mtDNA sequence against a database of 994 different mtDNA
sequences from modern humans.
Modern humans differed from each other in 8.0 $\pm$3.1 positions,
by contrast, the Neandertal genome had 27.0 $\pm$2.2 differences from
modern humans. 
In 1999, the same people successfully extracted a second mtDNA sequence
from the same Neandertal fossil. This study confirmed
the results of the first one, 
modern humans differed from each other by 10.9 $\pm$5.1 (range 1-35), the Neandertal
differed from humans by 35.3 $\pm$2.3 (range 29-43).
In 1999, they successfully extracted a mtDNA sequence from a
second Neandertal, a 29,000 year-old fossil (Mezmaiskaya).
The distance between Mezmaiskaya and a particular modern human sequence, known
as the reference sequence, was 22, compared to 27 for the first Neandertal
while the two Neandertals differed from each other in 12 positions.      
In 2000, scientists announced the sequencing of a third Neandertal mtDNA specimen
from a cave at Vindija, Croatia. The Neandertal
differed from modern humans by 34.9 $\pm$ 2.4 positions.

The conclusion was that the Neandertals lie at a statistically 
large distance from modern humans. 
Results in this paper suggest this conclusion being incorrect, 
in fact, the situation is quantitatively the same of fig. 2, where there
is a subpopulation whose distance from others is three or more times larger 
then the average distance and the most probable one.

Mungo Man, at variance with Neandertals, is an  anatomically modern man.  
The fossil, 60,000 years-old,
(older then the three Neandertal fossils)  was discovered in 1974 in the 
dry bed of Lake Mungo in New South Wales, Australia. 
Recently Mungo Man has attracted attention due to the 
extraction of mtDNA from fragments of his skeleton~\cite{Adcock}.
The authors identify differences between   
Mungo mtDNA and living aborigines mtDNA and conclude
that Mungo man belongs to a lineage diverging 
before the most recent common ancestor of contemporary humans. 
Also in this case, the argument is doubtful, in fact,  
as already discussed the rapid extinctions 
of mtDNA subpopulations are well evident in fig. 2.  

The conclusion (if any) of this work is that hardly mtDNA 
studies can be used to prove~\cite{Krings1997,Krings2000} 'Out of Africa' 
theory or disprove it~\cite{Adcock}. 
On the contrary, the study of DNA distribution in living population
allows for much more reliable results, especially if the study is performed on 
nuclear DNA which encodes information about all our ancestry.
Up to now, these studies mostly support 'Out of Africa' theory 
in its original form or in a more recent and less extreme one~\cite{Templeton}.

\bigskip
I thank Antonella Di Mattia,
Barbara Nelli and Michele Pasquini for many useful discussions and for a 
critical reading of the manuscript.

\newpage

\begin{figure}
\vspace{.2in}
\centerline{\psfig{figure=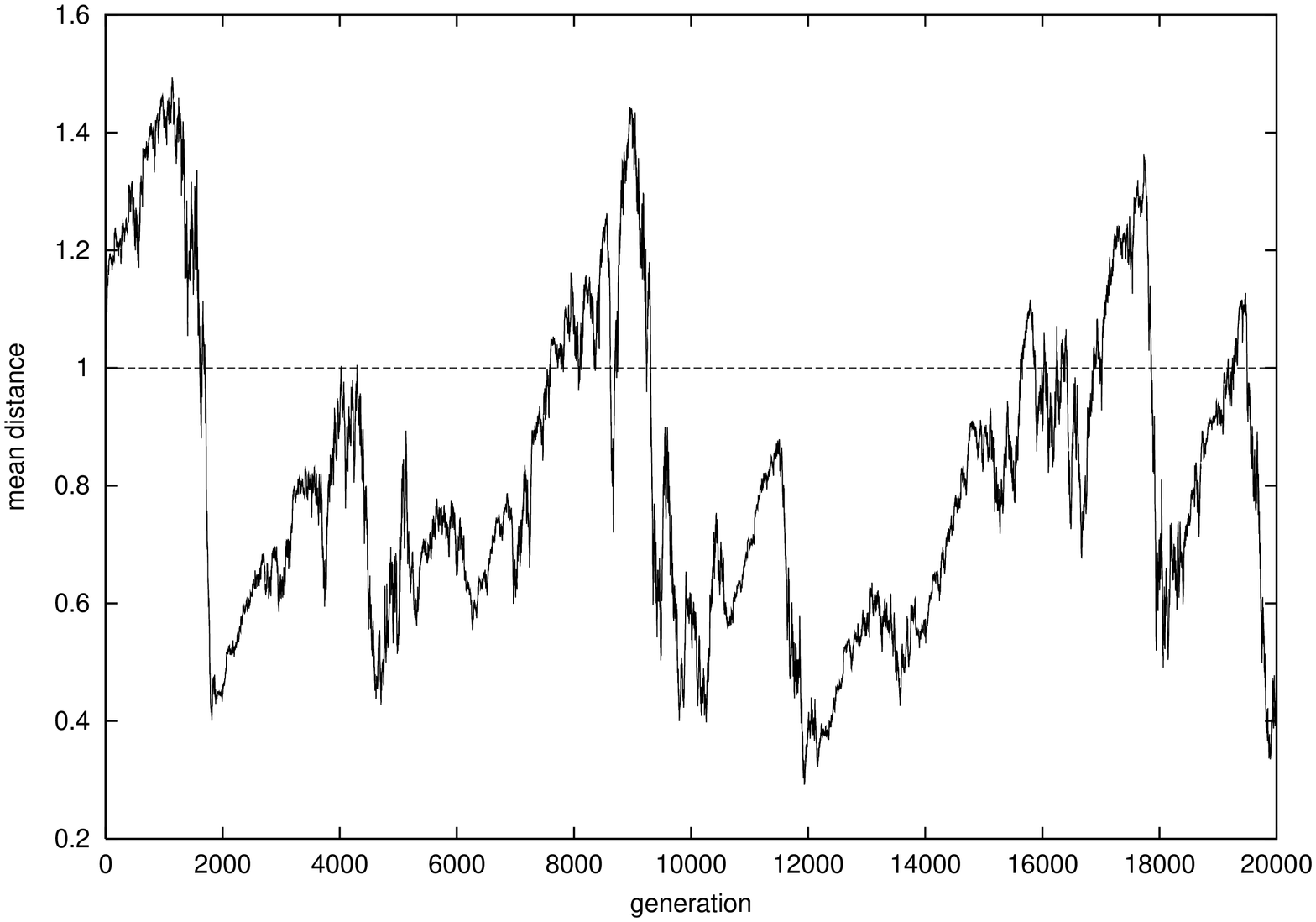,width=4.0truein,angle=270}}
\bigskip
\caption{ Mean distance of a single population individuals
as a function of time (generation).
The time evolution of the mean distance
is computed for $N=2000$, which is
sufficiently large to destroy all finite size effects.
This quantity remains random even in the infinite 
population limit.    
Notice that the mean distance is subject to abrupt
negative variations due to the extinction of large
subpopulations.}
\label{f1}
\end{figure}

\begin{figure}
\vspace{.2in}
\centerline{\psfig{figure=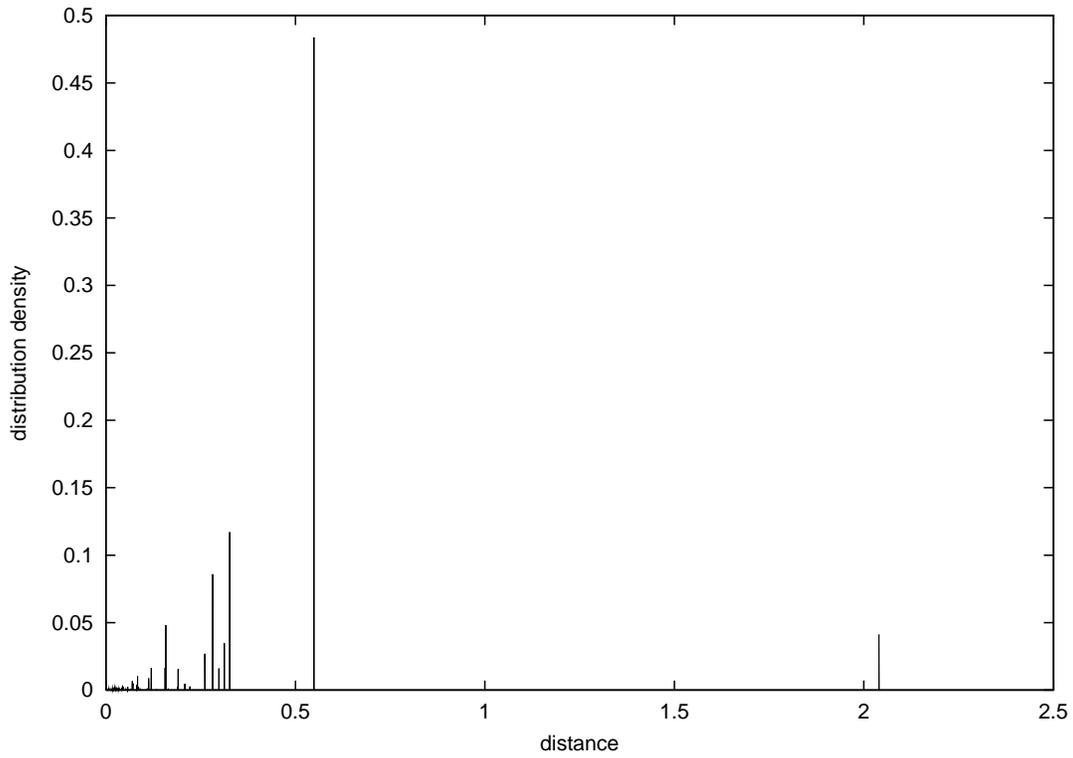,width=4.0truein,angle=270}}
\bigskip
\caption{Distribution density of distances in a single population.
Here we compute this quantity for a population of  2000 individuals.
The most important fact is that the distribution is very 
different from its process average $\exp(-x)$.
It is clear that individuals
naturally cluster in groups.
In fact, most of the distances assume a few of values 
corresponding to the distances between the major subpopulations.  
Notice that the largest distances are a few time larger than 
average distance and most probable one.}

\label{f2}
\end{figure}

\end{document}